\documentclass[%
 aip,
 apl,
 reprint,
 amsmath,amssymb
]{revtex4-1}

\usepackage{graphicx}
\usepackage{dcolumn}
\usepackage{bm}
\usepackage{gensymb}
\usepackage{siunitx}
\usepackage[english]{babel}

\newcommand{\wavenumber}[1]{\ensuremath{\SI{#1}{\per\centi\meter}}}
\newcommand{\mumetr}[1]{\ensuremath{\SI{#1}{\micro\meter}}}
\newcommand{\nmetr}[1]{\ensuremath{\SI{#1}{\nano\meter}}}

\begin{document}

\preprint{APS/123-QED}

\title{Controlling Nanoscale Air-Gaps for Critically Coupled Surface Polaritons by Means of Non-Invasive White-Light Interferometry.}

\author{Karsten Pufahl}\email{pufahl@tu-berlin.de}
 \affiliation{Institute of Optics and Atomic Physics, Technical University of Berlin, Strasse des 17. Juni 135, 10623 Berlin, Germany}
\author{Nikolai Christian Passler}\email{passler@fhi-berlin.mpg.de}
 \affiliation{Fritz-Haber-Institut der Max-Planck-Gesellschaft, Faradayweg 4-6,14195 Berlin, Germany}
 \author{Nicolai B. Grosse}
  \affiliation{Institute of Optics and Atomic Physics, Technical University of Berlin, Strasse des 17. Juni 135, 10623 Berlin, Germany}
\author{Martin Wolf}
 \affiliation{Fritz-Haber-Institut der Max-Planck-Gesellschaft, Faradayweg 4-6,14195 Berlin, Germany}
\author{Ulrike Woggon}
 \affiliation{Institute of Optics and Atomic Physics, Technical University of Berlin, Strasse des 17. Juni 135, 10623 Berlin, Germany}
\author{Alexander Paarmann}
 \affiliation{Fritz-Haber-Institut der Max-Planck-Gesellschaft, Faradayweg 4-6,14195 Berlin, Germany}

\date{\today}

\begin{abstract}
We report an experimental method to control large-area air-gaps in the nanometer range for evanescent coupling of light to surface waves such as surface plasmon polaritons or surface phonon polaritons.
With the help of spectrally resolved white-light interferometry we are able to stabilize and tune the gap with nanometer precision and high parallelism. Our technique is non-invasive, leaves the coupling area unobstructed, and the setup delivers reference-free real-time readout up to \mumetr{150} distance between the coupling prism and sample. Furthermore, we demonstrate the application  to prism coupled surface polariton excitation. The active gap control is used to determine the dispersion of a critically coupled surface phonon polariton over a wide spectral range in the mid infrared.
\end{abstract}

\maketitle


Numerous applications in the field of photonics rely on small and precisely controlled gaps between media of different refractive indices. 
In the vicinity of such interfaces, short-range evanescent waves exist that are inaccessible in the far-field. In the case of total internal reflectance, exponentially decaying waves leak into the lower index media.
Overlapping these near-fields allows light to tunnel through barriers that would otherwise reflect it. This near field coupling is the driving principle for various optical technologies such as fiber optic directional coupling\cite{Bergh1980},
near-field scanning optical microscopy\cite{Betzig1986}, total internal reflection fluorescence microscopy\cite{Axelrod1981}, and surface plasmon resonance spectroscopy\cite{Homola2003}.

As the evanescent decay of the light field occurs on the order of the wavelength or even below, nanometer precision control of the gap size is an essential requirement. While the technical difficulties have been solved for integrated photonics, evanescent coupling through large-area nanometer air-gaps is still challenging. Laterally extended coupling gaps are required when far-field light needs to be coupled to surface bound modes such as surface plasmon polaritons (SPP) or surface phonon polaritons (SPhP). Both polariton modes can also be excited with gratings\cite{Beaglehole1969,OHara2004}, nanostructures\cite{Caldwell2013, Wang2013} and tip-enhanced approaches\cite{Huber2005,Li2015}. However, evanescent prism coupling -- also known as attenuated total reflectance (ATR) -- is unsurpassed when well-defined k-vectors, broad-band coupling and minimal coupling loss is required.

The nature of separation by an air-gap makes such setups very prone to thermal expansion and vibrations with severe impact on the stability. 
Due to the lossy nature of both plasmons and phonons, depositing heat in the system is unavoidable, especially if combined with nonlinear optics. Large heat flux variations can also occur when the excitation beam is scanned over the polariton resonance. Thus a real-time feedback of the gap width is required to counter thermal expansion and stabilize the coupling conditions.

Nanometer resolving technologies such as glass scale encoders, capacitive sensors,  piezoresistive sensors, or differential transformers are readily available\cite{Fleming2013}. However, long-ranging surface waves require an unobstructed coupling region which prevents the placement of such sensors directly in the gap. Encoder solutions mounted at an offset position are subject to errors following from tilt and deformation which demand external compensation.

These technical challenges are likely the reason why coupling through an air gap, the so-called Otto geometry, has seen very little adaption since its invention\cite{Otto1968}. Most experiments make use of the Kretschman-Raether geometry where the sample is applied as a sub-wavelength thin film directly on the prism. Although inherently more robust to mechanical and thermal effects, this geometry is unsuitable for bulk or opaque samples and requires separate samples for each desired coupling strength.
Moreover, a precisely controlled coupling strength is important for the achievement of \emph{critical coupling} -- the point of loss-free coupling where intrinsic propagation losses match the radiative losses.
Using the Otto geometry is advantageous for systems where the propagation properties are not a priori known, and for the measurement of the entire polariton dispersion at critical coupling.

\begin{figure*}[t!]
\includegraphics[scale=0.5]{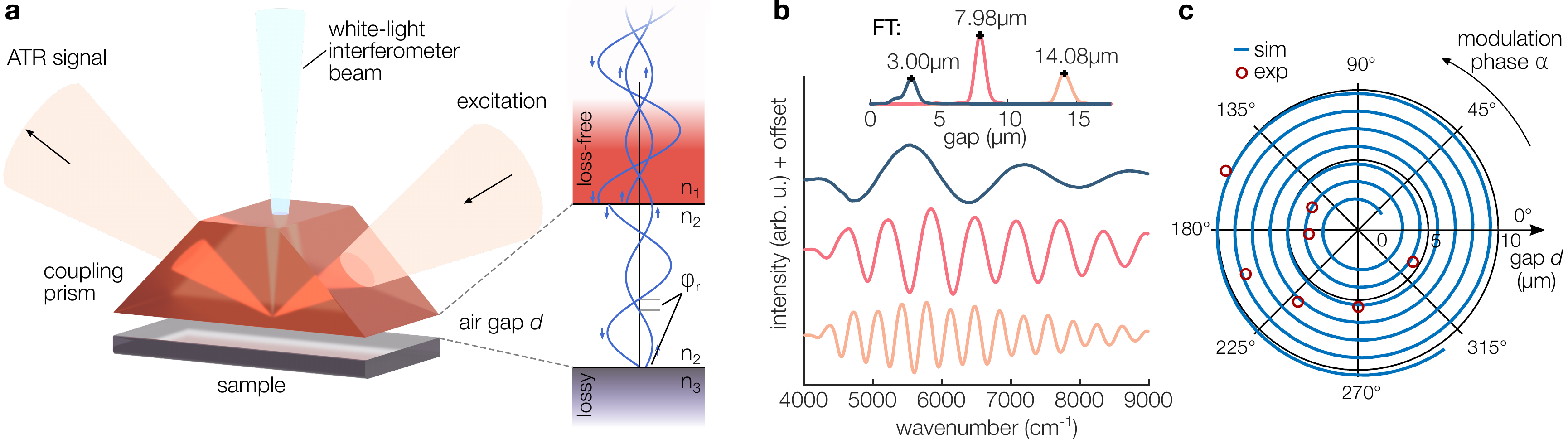}
	\caption{(a) Sketch of the ATR geometry with integrated white-light interferometry for high-accuracy absolute coupling gap control. Back reflections from interfaces interfere in the spectrometer and provide a unique spectral modulation.
(b) Reflectance spectra for an AlN sample (as in Fig. \ref{fgr:exp}). The Fourier transform of the signal allows to determine the coupling gap size $d$. (c) The modulation frequency plotted together with its phase. In this representation, changing the coupling gap describes a spiral path. Concurrent evaluation of both quantities improves the accuracy and serves as a verification.}
  \label{fgr:sketch}
\end{figure*}

In this work we present a simple technique that reduces the mechanical complexity and allows sensing the air-gap in a direct and non-invasive way which is highly desired to harness the full potential of air-gap evanescent coupling. We employ spectrally resolved white-light interferometry to reach arcsecond parallelism and nanometer absolute distances for the coupling gap. Our method grants remote in-situ monitoring of the coupling distance without perturbing the surface polariton propagation\cite{Passler2017,Grosse2018} and allows real-time feedback to stabilize thermal drifts.
The measured signal is a unique function of the gap-size and initial referencing is not required compared to monochromatic interferometers.

We have implemented two experimental setups based on distinct requirements for the excitation of SPPs and SPhPs. For SPPs in the visible (VIS) and near infrared (NIR), a quartz coupling prism is employed, while for SPhPs in the mid IR, the prism material is thallium bromo-iodide (KRS5). The light source used for the interferometry is chosen to be compatible with the respective prism transmission window, being a white LED source (Cree XPEWHT-L1, $4000~K$, $430-730~nm$) for the quartz prism, and a tungsten NIR source (Thorlabs SLS201L, $2796~K$, $360-2600~nm$) for the KRS5 prism. 

As sketched in Fig. \ref{fgr:sketch} (a), the top of a conventional isosceles triangular prism is grounded off parallel to the probing interface. A fiber coupled white-light beam is fed through the prism top and focused onto the sample with $1 \degree$ opening angle. Reflections from the prism-air interface and the air-sample interface are fed back to the fiber coupler. The divergence of the white-light is small enough such that the Rayleigh length is much larger than the coupling gap and both contributions can interfere in the far-field.
For detection of the interference signal, the back-reflected signal is passed through a beam splitter to separate the forward and backward propagating signals.
A significant part of the light is reflected back from the fiber end due to coupling losses. This can be counteracted by placing a pair of crossed polarizers in the beam path, ensuring that only light coming out of the non-polarization-maintaining fiber can reach the spectrometer.
Alternatively, a bifurcated fiber bundle has been used to split the in and outgoing white-light (Thorlabs RP21). The white-light signal is analyzed in a UV/VIS spectrometer (Thorlabs CCS200) for the quartz prism setup, and an InGaAs array based spectrometer (Ocean Optics NIRQuest512) for the KRS5 prism, respectively.

The light for polariton excitation, on the other hand, is incident and reflected at the bottom of the prism at angles above total internal reflection. Due to this angular separation, the gap width can be measured simultaneously to the reflectance spectrum featuring the polariton resonances (see Fig. \ref{fgr:sketch}(a)). 

The detectable range for the distance readout depends on the signal-to-noise ratio and the wavelength resolution of the spectrometer. The former limits the lower end of the range where the modulation period becomes wider than the spectrum. The latter becomes the limiting factor when the modulation period approaches the spectrometer resolution, which happens at the upper end of the range. For the presented VIS setup, detectable gaps span almost 3 orders of magnitude between \nmetr{370} and \mumetr{150}, while the NIR setup covers a range from \nmetr{500} up to \mumetr{60}.
For small gap sizes precise angular adjustment is mandatory. For sample sizes in the millimeter range angular adjustment of a few arcsecond is required to prevent the collision of sample and prism.

\begin{figure*}[t]
\includegraphics[width=\textwidth]{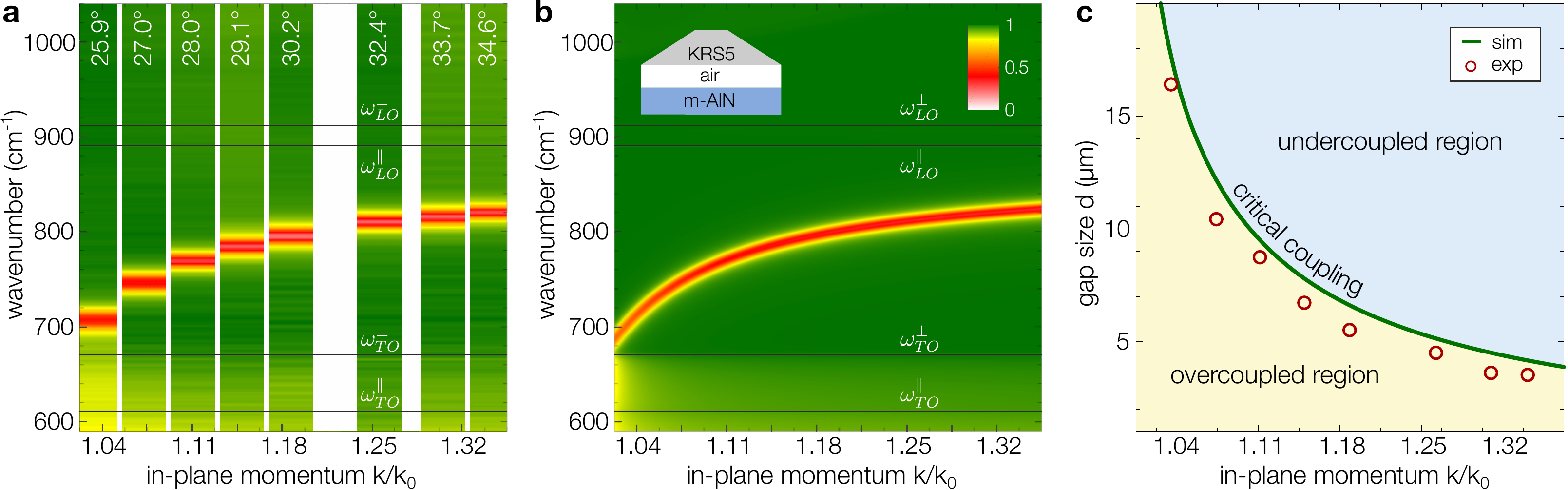}
  \caption{Excitation of a SPhP in AlN at critical coupling conditions. (a) Experimental reflectance data at various incidence angles (i.e. in-plane momenta) exhibiting a deep reflectance dip due to excitation of the SPhP. For each angular step the gap was adjusted to the critical coupling gap $d_{\text{crit}}$. Optimal coupling was achieved across the entire dispersion curve. (b) Simulated reflectance map featuring the same single SPhP dispersion of AlN, being in excellent agreement with the experimental data. The simulations were performed employ a $4\times 4$ transfer matrix approach\cite{Passler2017a} in order to account for the anisotropy of m-AlN, see Supplementary Material for details. (c) Theoretical and experimental critical gap $d_{\text{crit}}$ as a function of in-plane momentum, i.e. along the SPhP dispersion. Due to the strong variation of $d_{\text{crit}}$, the precise control and readout of the gap $d$ via white-light interferometry is highly desirable for the correct mapping out of polariton dispersions.}
  \label{fgr:exp}
\end{figure*}

The back reflected white-light signal carries a distance dependent spectral modulation. The actual gap width $d$ can be calculated from this unambiguous signature.  The condition for constructive interference from the two surfaces is 
\begin{align}
d\cdot n_2\cdot\cos \theta=(m+\frac{\varphi_\text{r}}{2\pi})\cdot\lambda,\quad m=1,2,3\dots,
\end{align}
where $d$ is the interface spacing, $n_2$ the refractive index in the gap, $\theta$ the tilt angle of the white-light beam with respect to the sample surface normal, and $\varphi_\text{r}$ represents the phase retardation due to reflection at the sample surface. Light reflected at the prism surface is in-phase with the incident wave ($n_1 > n_2$ as required by ATR spectroscopy). At the sample surface, however, the phase retardation depends on the absorption in the medium and can be between $0\degree$ and $180\degree$, as illustrated in Fig. \ref{fgr:sketch} (a). Since metals supporting SPPs are absorptive in the spectral region of the white-light source, knowing $\varphi_\text{r}$ is essential, while in non-absorptive dielectrics supporting SPhPs the phase retardation can be neglected. Although the influence of $\varphi_\text{r}$ does not alter the periodicity of the signal with distance, it does give an offset which is important for the absolute distance. Disregarding $\varphi_\text{r}$ is likely the reason why previous experiments on the Otto ATR geometry have underestimated the critical coupling distance\cite{Falge1973, Futamata1994}. Consequently, the reflection phase has to be known or measured by ellipsometry prior to calculating the gap from the reflectance spectrum.
In Fig. \ref{fgr:sketch} (b), exemplary spectra are shown after referencing the data to a spectrum taken at a gap width $d\gg \mumetr{150}$, i.e. far above the resolvable coherence length.

The calculation of the gap width $d$ from the spectral data can be performed with a Fourier transformation. For this means, the peak position $f$ in the Fourier spectrum is evaluated as follows:
\begin{align}
2d\cdot n_2 \cdot \cos\theta = f + \frac{\partial\varphi_\text{r}(k)}{\partial k}
\label{eqn:modulation}
\end{align}
Where the left hand side is the optical path length of the light in the gap and the right hand side is composed of two influences on the spectral modulation pattern: Firstly, the modulation frequency $f$ resulting from spacing of the two interfaces, and secondly the dispersion of the reflection phase $\varphi_\text{r}$, being non-zero for absorbing materials where resonances alter the reflection phase.
There are two ways to handle this dispersion when computing $d$ by means of a Fourier transform. If a material resonance lies outside the spectral window and the phase is relatively flat, it can be approximated with a linear slope, which gives a constant offset to the measurement ($d_\text{offset} = \frac{1}{2n_2 \cos(\theta)}\frac{\Delta\varphi_\text{r}}{\Delta k}$). For materials with a nonlinear reflection phase or multiple resonances inside the reflectivity spectrum it is suggested that the phase is precompensated before applying the Fourier transform. This can be achieved by multiplying the modulation signal with $\exp\{-i(\pi-\varphi_\text{r}(k))\}$.

Additional improvement is gained by evaluating not only the frequency but also the phase $\alpha$ of the modulation. Fig. \ref{fgr:sketch} (c) shows a polar plot with the frequency on the radial axes and the corresponding phase in azimuthal direction. In this representation, a changing gap leads to a spiral around the origin. Especially for very narrow gaps with few or even single cycle modulation, evaluation of the phase significantly improves the accuracy.

Besides absolute gap width readout, our method allows for parallel alignment of prism and sample with high accuracy ($\pm 1 ''$). This is achieved by firstly optimizing the contrast of the interference signal, and secondly by laterally scanning the white-light measurement spot over the sample. By this means, the obtained constant gap width assures constant polariton excitation conditions across the complete light spot, being crucial for obtaining well-defined polariton resonances in the ATR spectra.


In order to demonstrate our method, we present a measurement of critically coupled SPhPs excited in m-cut hexagonal aluminum nitride (AlN) in the KRS5 prism setup. In Fig. \ref{fgr:exp} (a), the reflectance data taken at 8 different angles of incidence $\theta$ is plotted, where the dips in the reflectance (red spots) correspond to the excitation of a SPhP. Clearly, this excitation frequency shifts with varying in-plane momentum $k/k_0=n_{\text{KRS5}} \sin \theta$ ($n_{\text{KRS5}}\approx 2.4$), following the characteristic shape of the SPhP dispersion at a single interface\cite{Raether1988}. Fig. \ref{fgr:exp} (b) shows the corresponding theoretical reflectance signal that was calculated with an anisotropy-aware transfer matrix method\cite{Passler2017a}. The SPhP dispersion branch which is visible between $\wavenumber{700}$ and $\wavenumber{850}$ is in excellent agreement with our experimental results. We note that while the excitation of a SPhP at critical coupling conditions ideally leads to zero reflectance, we here observe dips with non-zero minimum reflectance due to the finite spectral width $\Delta\tilde{\nu}= \wavenumber{5.7}$ of our excitation source (see Supplementary Information for details).

The air gap of critical coupling conditions $d_{\text{crit}}$, i.e. where the SPhP is excited most efficiently resulting in minimized reflectance\cite{Passler2017}, strongly depends on the in-plane momentum, as shown in Fig. \ref{fgr:exp} (c) (green solid line). Therefore, the air gap has to be adjusted in the experiment at each incidence angle individually, optimizing for the deepest dip in the reflectance. These experimental critical gaps $d_{\text{crit}}$ are plotted as red circles in Fig. \ref{fgr:exp} (c), following the theoretically calculated curve (green solid line). The origin of the systematic offset is explained in the Supplementary Information. Note that for deviations from the curve of critical coupling, the SPhP excitation looses efficiency either because of large radiative losses into the prism at close gaps (overcoupled region), or because of insufficient overlap with the evanescent wave from the prism at large gaps (undercoupled region).

The critical gap $d_{\text{crit}}$ markes the point where the radiative losses equal the intrinsic losses\cite{Neuner2009} and is hence a function of the propagation lifetime of the SPhP.
It can vary strongly for different materials or even for different polaritons in a single system\cite{Passler2017a}. Therefore, it is essential to have precise control over the gap in order to excite polaritons under critical coupling conditions and to map out their dispersion curves.
Measurements with air gaps down to \nmetr{370} -- as required for surface plasmon polariton excitation -- are shown in the Supplementary Information, featuring the quartz prism setup.


In summary, the feasibility of white-light-based interferometry for surface polariton excitations was demonstrated experimentally with an m-cut hexagonal AlN sample. The method has proven to be a useful tool for the precise control over the evanescent coupling strength by providing live feedback of the coupling gap without disturbing the polariton excitation. The gained control over the air gap was employed to achieve critical coupling conditions at any in-plane momentum, hence allowing to experimentally reconstruct the SPhP dispersion of the AlN sample. The results were verified with theoretical calculations showing an excellent agreement.

The resolution limit of our method is determined by external parameters like the spectrometer resolution, the wavelength calibration, and ellipsometry data of the materials involved, and is therefore further extendable.
Given that the reflectivity spectrum can be measured with high precision, the dispersion of the reflection phase is left as the largest source of error when determining $d$. Here, the accuracy of the ellipsometry data is important, when single digit nanometer resolution is required. As a guide we have estimated $d$ for a Au surface in the visible region by comparing available ellipsometry data\cite{Johnson1972,Palik1998,Gao2012,Babar2015}. This has led to an uncertainty of $\Delta d = \nmetr{1.7}$. In comparison, Au film samples that were coated in the same batch and measured individually with spectral ellipsometry resulted in an error of less than \nmetr{0.2}. Completely disregarding the influence of the reflection phase dispersion would result in an error as large as \nmetr{170} for Au.

The range of \nmetr{370} to \mumetr{150} as obtained in our setup should cover the majority of applications for both SPPs and SPhPs. Another advantage of our method is the spatial and galvanic isolation provided by the fiber connection between the sensing head and the detection system, which enables functionality under a broad range of measurement conditions. Furthermore, applications in nonlinear optics can benefit from the real-time feedback to stabilize the coupling even under strong temperature drifts, as happens when tuning the frequency
from off-resonance to on-resonance excitation of the polariton. Finally we emphasize that the presented method is not limited to surface polariton propagation but has various other potential applications where well-defined air-gap based evanescent coupling is required. Among others, this includes ATR-spectroscopy, waveguide coupling or excitation of ring-resonator structures.

In conclusion, we have presented a method to overcome technical challenges in air-coupled Otto geometry setups. Our technique allows angle resolved spectroscopy with constant critical coupling conditions, thus granting the possibility to experimentally determine polariton dispersions of any flat sample. Furthermore, tuning the coupling gap allows critical coupling for a large wavelength range compared with systems that are optimized for a fixed wavelength such as grating couplers or Kretschmann geometries. Quantifying the coupling gap not only assures repeatability and comparability but can also serve to measure propagation losses from previously unknown or uncharacterized samples. By providing the necessary control over polariton coupling with tunable air gap in a reliable and widely applicable manner, we hence envision our method to pave the way for multiple future studies in the field of polaritonic nanophotonics.

\section{Supplementary Material}
Details on the experimental procedure and the free-electron laser (S1), details on the fitting procedure for the ATR data (S2), and details of the quartz prism setup for VIS and NIR excitation (S3).

\section{Acknowledgement}
This work was supported by the German Research Foundation under grant WO 477/35-1. The authors thank J.D. Caldwell (Vanderbilt University) for providing the AlN sample.

\bibliography{whitelight}
\end{document}


\title{Supplementary Information: \\ Controlling nanoscale air-gaps for critically coupled surface polaritons by means of non-invasive white-light interferometry}

\author{Karsten Pufahl}\email{pufahl@tu-berlin.de}
 \affiliation{Institute of Optics and Atomic Physics, Technical University of Berlin, Strasse des 17. Juni 135, 10623 Berlin, Germany}
\author{Nikolai Christian Passler}\email{passler@fhi-berlin.mpg.de}
 \affiliation{Fritz-Haber-Institut der Max-Planck-Gesellschaft, Faradayweg 4-6,14195 Berlin, Germany}
 \author{Nicolai B. Grosse}
  \affiliation{Institute of Optics and Atomic Physics, Technical University of Berlin, Strasse des 17. Juni 135, 10623 Berlin, Germany}
\author{Martin Wolf}
 \affiliation{Fritz-Haber-Institut der Max-Planck-Gesellschaft, Faradayweg 4-6,14195 Berlin, Germany}
\author{Ulrike Woggon}
 \affiliation{Institute of Optics and Atomic Physics, Technical University of Berlin, Strasse des 17. Juni 135, 10623 Berlin, Germany}
\author{Alexander Paarmann}
 \affiliation{Fritz-Haber-Institut der Max-Planck-Gesellschaft, Faradayweg 4-6,14195 Berlin, Germany}

\date{\today}
\maketitle
\beginsupplement

\section{Experimental}

The experimental setup is sketched in Fig. 1 of the main text. The Otto geometry employed for the NIR measurements uses three motorized actuators (Newport TRA12PPD) to control the position of the triangular coupling prism (KRS5, 25mm high, 25mm wide, angles of \dg{30}, \dg{30}, and \dg{120}, with the \dg{120} edge cut-off, Korth Kristalle GmbH) relative to the AlN sample. The actuators allow to control the air gap $d_{gap}$ continuously by pushing the prism against springs away from the sample. The \dg{120} prism edge is ground-off parallel to the backside in order to perpendicularly couple in the light for the gap measurement via white-light interferometry. 

At external incidence angles $\theta_{ext}$ other than normal to the prism side ($\dg{30}$), the refraction due to the high refractive index of KRS5 ($n_{\text{KRS5}}\approx 2.4$) needs to be taken into account.The internal incidence angle $\theta$ inside the prism is $\theta=\dg{30}+\arcsin \left( \sin \left( \theta_{ext}-\dg{30} \right) / n_{\text{KRS5}} \right)$. In this configuration, angles from below the critical angle of total internal reflection ($\theta_{crit}\approx \dg{25}$) up to about \dg{52} are accessible. Different in-plane momenta can be accessed via the incidence angle $\theta$ by rotating the entire Otto geometry, thus allowing for mapping out the complete dispersion curve of the AlN SPhP experimentally.

In the MIR setup, we employ a mid-infrared free electron laser (FEL) as tunable, narrowband, and p-polarized excitation source. Details on the FEL are given elsewhere\cite{Schollkopf2015}. In short, the electron gun is operated at a micropulse repetition rate of $1$ GHz with a macropulse duration of \musec{10} and a repetition rate of $10$ Hz. The electron energy is set to 26 MeV, allowing to tune the FEL output wavelength between $\sim \mumetr{9-22}$ ($\sim \wavenumber{1100-450}$, covering the spectral range of the AlN reststrahlen band) using the motorized undulator gap, providing ps-pulses with typical full-width-at-half-maximum (FWHM) of $<\wavenumber{5}$. The macropulse and micropulse energies are $\sim30$ mJ and $\sim10~\text{\miu J}$, respectively.

The signal reflected off the prism hypotenuse is detected with a pyroelectric sensor (Eltec 420M7). A second sensor is employed to measure the power reflected off a thin KBr plate placed prior to the setup in the beam. This signal is used as reference for normalization of all reflectivity signals, and is measured on a single shot basis in order to minimize the impact of shot-to-shot fluctuations of the FEL. The spectral response function of the setup is determined by measuring the reflectivity at large gap widths $d_{gap}\approx \mumetr{60}$, where the prism features total internal reflection across the full spectral range. This is done for each incidence angle individually, and all spectra are normalized to their respective response function. 

The reflectivity spectra shown in Fig. 3a of the main text are measured at gaps of critical coupling conditions $d_{crit}$, in order to acquire the dispersion curve of the AlN SPhP. Experimentally, $d_{crit}$ is found by optimizing for the deepest dip in the reflectivity at the SPhP resonance frequency for each incidence angle individually. However, the FEL output center frequency fluctuates in time in a range of about $\pm \wavenumber{2}$ due to shifts in the electron energy, making it difficult to optimize for a narrow reflectivity dip. As the SPhP dip is broader in the overcoupled than in the undercoupled region, this experimental difficulty is likely the cause of the constant offset towards the overcoupled region of the experimental critical gaps compared to the theoretical curve, see Fig. 3c of the main text.

\section{Fitting Procedure}
The sample is an m-cut hexagonal AlN crystal, meaning that the extraordinary c-axis lies in the surface plane. In such a crystal orientation, the anisotropy has a crucial impact on the SPhP dispersion and has to be fully considered for the simulation of the dielectric response of the sample. For this means we employ a $4 \times 4$ transfer matrix formalism that has been specifically developed for a multilayer system of highly dispersive media, where each material can either be isotropic or fully anisotropic\cite{Passler2017a}. 

All spectra composing the experimental dispersion curve (see Fig. 3a) were fitted simultaneously in a global fitting procedure. The gaps were fixed to the experimental values, while the incidence angles and a constant amplitude factor were fitted individually for each spectrum. On the other hand, the spectral width of the FEL ($\Delta\nu=\wavenumber{5.65}$), the angle spread ($\Delta\theta=\dg{0.24}$), the ordinary (o) and extraordinary (e) phonon frequencies and the damping $\gamma$ of AlN ($\nu_{TO}^{o}=\wavenumber{672.3}$, $\nu_{TO}^{e}=\wavenumber{612.5}$, $\nu_{LO}^{o}=\wavenumber{906.1}$, $\nu_{LO}^{e}=\wavenumber{878.1}$, $\gamma=\wavenumber{3.0}$), and the azimuth angle of the crystal ($\phi = \dg{48.4}$) were fitted globally. Note that the azimuth rotation angle $\phi$ of the AlN sample was not known prior to the experiments, but could easily be extracted by the fitting procedure. 

The resultant values listed above were then used to numerically calculate the critical gap $d_{crit}$ (green curve in Fig. 3c of the main text) as a function of the in-plane momentum $k_{\parallel}=n_{\text{KRS5}} \sin \theta$, by determining the gaps were the reflectivity dip of the SPhP reaches its minimum. Finally, the reflectivity map exhibiting the AlN SPhP shown in Fig. 3b of the main text was calculated at the critical gap $d_{crit}(\theta)$ calculated beforehand, revealing the entire SPhP dispersion in the rotated m-cut AlN crystal.

\section{Quartz Prism Setup}
\begin{figure*}[t]
\includegraphics[scale=1]{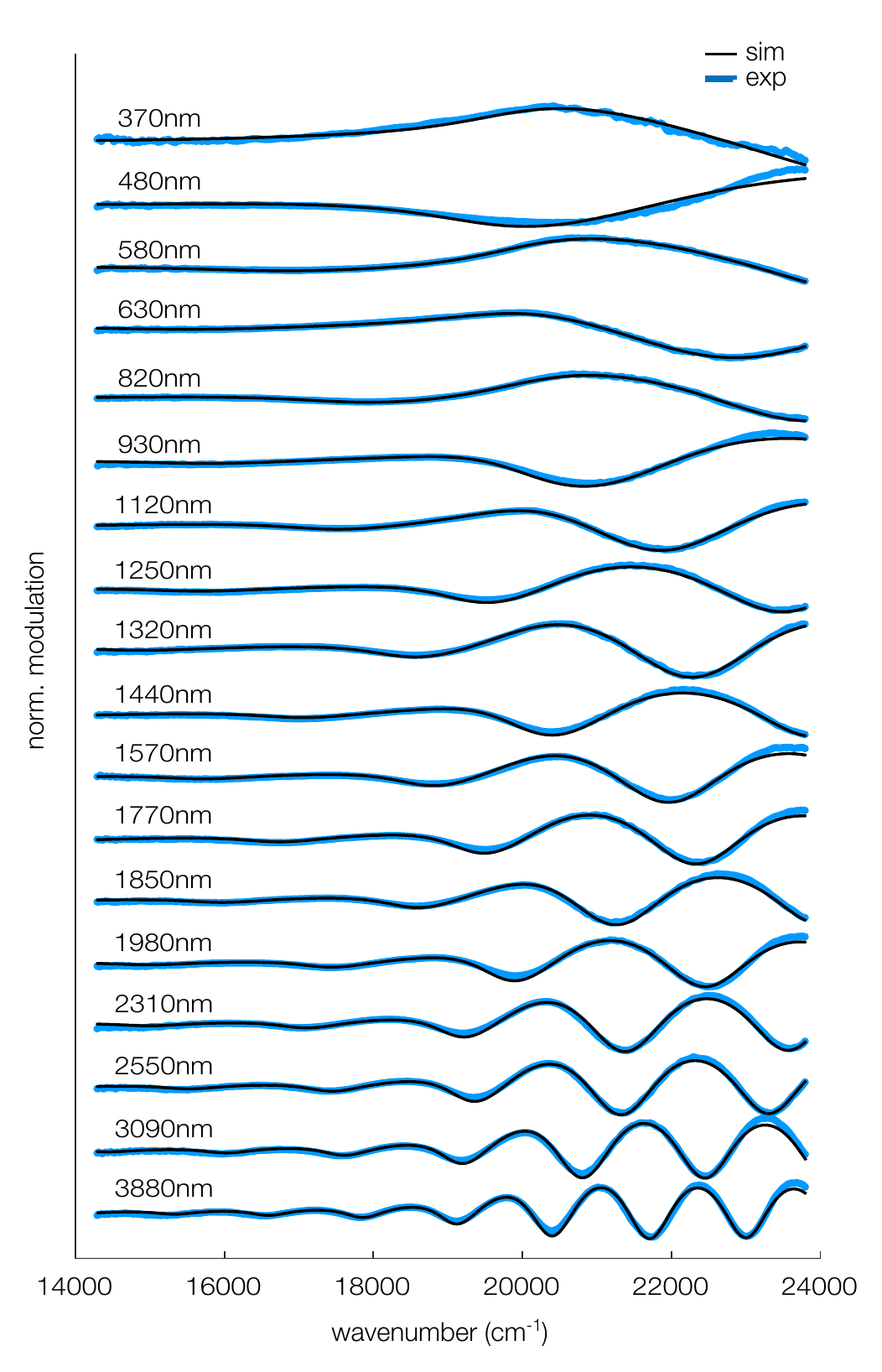}
  \caption{
  Series of gap size measurements during excitation of surface plasmon polaritons on a Au surface using the quartz prism Otto geometry setup. The spectra match well with the fitted model from which the gap size was extracted. The modulation depth is significantly higher at the blue end of the spectra where the prism-air and the air-sample reflections are more balanced than towards the NIR.
  }
  \label{fgr:spp_distances}
\end{figure*}

Critical coupling gaps for the VIS and NIR excitation of surface plasmon polaritons are an order of magnitude below those of surface phonon polaritons. In the quartz prism setup, illumination with shorter wavelength (VIS range) allows to detect gaps well below \mumetr{1}.
Figure \ref{fgr:spp_distances} shows an exemplary series of coupling gaps for surface plasmon polariton excitation on a Au surface. The characteristic inter-band transitions of Au absorb a large portion of the incident whitelight -- in particular at the very blue end of the spectrum. This is beneficial for the modulation contrast, as the reflectivity of the prism-air interface and the air-sample interface are more balanced.
The required parallelism to avoid collision of prism and sample was 
estimated from the sample size and the coupling gap. Approaching the 
surface up to 370 nm requires at least 1.9 arcsecond angular precision, 
which is easily achieved in our setup.
To achieve this parallelism, the sample was placed in the pivot point of a dual goniometer.
The probe beam was scanned normal to the prism surface (ground, $\lambda/10$ flatness) and the angular adjustment was carried out until until no shift of the modulation pattern was noticeable. While for very small gaps the spectra become flat 
and partially ambiguous, gap sizes down to 200 nm are readily achievable.


\bibliography{supplMat}